\documentstyle[12pt]{article}
\input epsf

\def\U1{$U(1)$}
\def\SU5{$SU(5)$}
\def\SO10{$SO(10)$}
\def\422{$SU(4)\otimes SU(2)_L \otimes SU(2)_R$}

\def\diag.{\hbox{diag.}}

\def\M_U{\hbox{$M_U$}\ }
\def\M_P{\hbox{$M_P$}\ }

\def\MSSM+N{\hbox{MSSM+$\nu$}}

\newcommand\beq{\begin{equation}}
\newcommand\eeq{\end{equation}}
\newcommand\bea{\begin{eqnarray}}
\newcommand\eea{\end{eqnarray}}
\newcommand\ba{\begin{array}}
\newcommand\ea{\end{array}}

\begin{document}
\baselineskip 24pt
\newcommand{\sheptitle}
{Lepton Flavour Violation in the Constrained MSSM with Natural
Neutrino Mass Hierarchy}

\newcommand{\shepauthor}
{T. Bla\v{z}ek$^*$ and S. F. King}

\newcommand{\shepaddress}
{Department of Physics and Astronomy, University of Southampton \\
        Southampton, SO17 1BJ, U.K}

%%%%%%%%%%%%%%%%%%%%%%%%%%%%%%%%%%%%%%%%%%%%%%%%%%%%%%%%%%%%%%%%%%%%%%%%%%%%%%%
%                                  ABSTRACT
%%%%%%%%%%%%%%%%%%%%%%%%%%%%%%%%%%%%%%%%%%%%%%%%%%%%%%%%%%%%%%%%%%%%%%%%%%%%%%%
\newcommand{\shepabstract}
{We present predictions for $\mu \rightarrow e \gamma$ and 
$\tau \rightarrow \mu  \gamma$ in the CMSSM in which there is a
natural hierarchy of neutrino masses resulting from the sequential
dominance of three right-handed neutrinos.
We perform a global analysis of this model in the
$(m_0,M_{1/2})$ plane, assuming
radiative electroweak symmetry breaking, and including
all observed laboratory data.
We confirm that a large (small) $\tau \rightarrow \mu \gamma$
rate results from the dominant right-handed neutrino being the heaviest
(lightest) one. We show that the $\mu \rightarrow e \gamma$ rate
may determine the order of the sub-dominant 
neutrino Yukawa couplings in the flavour basis, but may also be sensitive
to effects beyond the leading log and leading mass insertion approximations.
We also show that
the $\mu \rightarrow e \gamma$ rate is independent of $\theta_{13}$,
but measurement of this angle may determine a ratio of sub-dominant
Yukawa couplings.}

\begin{titlepage}
\begin{flushright}
hep-ph/0211368
\end{flushright}
\begin{center}
{\large{\bf \sheptitle}}
\\ \shepauthor \\ \mbox{} \\ {\it \shepaddress} \\ 
{\bf Abstract} \bigskip \end{center} \setcounter{page}{0}
\shepabstract
\begin{flushleft}
\today
\end{flushleft}

\vskip 0.1in
\noindent
$^*${\footnotesize On leave of absence from 
the Dept. of Theoretical Physics, Comenius Univ., Bratislava, Slovakia}

\end{titlepage}

\newpage

\section{Introduction}

Atmospheric and solar neutrino experiments have provided convincing
evidence for neutrino masses and mixings. The current minimal best fit
interpretation of the data involves three light neutrinos with
\cite{talks}
\bea
|\Delta m_{32}^2| & = & (1.7-3.3)10^{-3}{\rm eV^2}, \ \
\sin^22\theta_{23} =   0.93-1.0, 
\label{neutrinodata1}\\ 
\Delta m_{21}^2 & = & (4-10)10^{-5}{\rm eV^2}, \ \
\sin^22\theta_{12} =  0.71-0.89,
\label{neutrinodata}
\eea
where $m_{ij}^2=m_i^2-m_j^2$ are the neutrino mass squared differences
and the approximate 1$\sigma$ ranges are shown. 
These masses and mixings represent the first solid evidence of
new physics beyond the Standard Model. 

The most elegant explanation of small neutrino masses continues
to be the see-saw mechanism \cite{seesaw,Mohapatra:1979ia}. 
According to the see-saw
mechanism, lepton number is broken at high energies due to 
right-handed neutrino Majorana masses, resulting in small left-handed
neutrino Majorana masses suppressed by the heavy mass scale.
A natural implementation of the see-saw mechanism which can account
for a neutrino mass hierarchy and a large atmospheric mixing angle
is single right-handed neutrino dominance \cite{King:1998jw}.
See also \cite{Barbieri:1999pe,Altarelli:1999dg}.
It is also possible to account for a
large solar neutrino angle in this framework
\cite{King:1999mb}. For example a large solar angle may result from
sequential right-handed neutrino dominance due to 
the leading sub-dominance of a second
right-handed neutrino \cite{King:1999mb}, leading to a full neutrino mass
hierarchy $m_1\ll m_2 \ll m_3$. 

In charged lepton interactions of the Standard Model
the lepton flavour violation (LFV) is suppressed
by the heavy see-saw mass scale, or equivalently the small left-handed
Majorana masses, and is practically unobservable. However in 
low energy supersymmetric
(SUSY) theories the situation dramatically changes for the better.
In SUSY the LFV is imprinted on the
slepton mass matrices, resulting in slepton masses which are
off-diagonal in flavour space. The off-diagonal slepton masses
can then induce LFV in loops suppressed only by the SUSY
breaking scale, multiplied by ratios of the off-diagonal slepton mass to the
diagonal slepton mass 
\cite{Borzumati:1986qx,Gabbiani:1996hi,Hisano:1995cp,King:1998nv}.

The experimental prospects for improving
the limits or actually measuring LFV processes are very promising.
The 90\% C.L. limits of 
${\rm BR}(\tau \rightarrow \mu \gamma)< 1.1\times 10^{-6}$ 
\cite{Ahmed:1999gh}
and ${\rm BR}(\mu \rightarrow e \gamma)< 1.2\times 10^{-11}$
\cite{Brooks:1999pu}
are particularly stringent in constraining SUSY models.
In fact, these limits will be lowered in the next 2-3 years
as the present B factories,
inevitably producing tau leptons along with the b quarks, will collect
15-20 times more data and as the new 
$\mu\to e\gamma$ experiment at PSI probes the branching ratio 
down to $10^{-14}$
\cite{BABAR_tmg:2002, meg_at_PSI:2002}.
The forthcoming experimental results are clearly a powerful
incentive for theoretical studies of 
$\tau \rightarrow \mu \gamma$ and $\mu \rightarrow e \gamma$,
and this provides an important underlying motivation for the present
paper. It should be noted however that LFV violation can originate from 
other effects unrelated to the see-saw mechanism, for example
colour triplet scalars in GUT models, or off-diagonal slepton
masses generated from flavour violating effects in the
SUSY breaking sector.
In any observation of LFV, such effects would need to be disentangled
from the effects due to the see-saw mechanism in the CMSSM considered here.

There is a huge literature on LFV in the framework of the CMSSM, where 
the constraint of universal scalar mass $m_0$ and trilinear mass
$A_0$ implies that the slepton mass non-universality originates
entirely from RG effects due to right-handed neutrinos
between the GUT scale and the lightest right-handed neutrino mass scale
\cite{huge}. There is a much smaller literature concerned with LFV in 
models based on single right-handed neutrino dominance
\cite{Blazek:2001zm,Lavignac:2001vp}.
In our previous paper \cite{Blazek:2001zm} we made the important observation 
that a large $\tau \rightarrow \mu \gamma$ rate results from models in which 
the dominant right-handed neutrino is the heaviest one,
and the neutrino Yukawa matrix has a lop-sided form with an order unity
Yukawa coupling in the 23 position of the matrix
\[
   Y^{\nu} \approx
            \left( \begin{array}{ccc}
                    0 & 0 & 0    \\
                    0 & 0 & 1    \\
                    0 & 0 & 1
                         \end{array}
                                    \right). 
\]
In subsequent papers \cite{Lavignac:2001vp} a bottom-up analysis of 
various types of right-handed neutrino dominance was
discussed and $\tau \rightarrow \mu \gamma$ and
$\mu \rightarrow e \gamma$ were both considered in terms
of coefficients $C_{ij}$ which parametrise the off-diagonal slepton
masses to leading log approximation in the mass insertion approach to LFV.
Bounds on the coefficients $C_{ij}$ 
were presented as a function of the sneutrino and second gaugino
masses $m_{\tilde{\nu}}$ and $M_2$,
and $C_{ij}$ were related to Yukawa couplings and right-handed
neutrino masses for 
each of the different types of right-handed neutrino dominance 
\cite{Lavignac:2001vp}.

In the present paper we provide a dedicated analysis of 
a particular class of right-handed neutrino dominance models, namely 
sequential dominance (SD). The reason why we choose to focus on SD
is that it provides a particularly elegant application 
of the see-saw mechanism to the LMA MSW solution, and these
are the only models where a full neutrino mass hierarchy
$m_1\ll m_2 \ll m_3$ arises naturally. 
Another attractive feature of SD models is that
it may become possible to relate
the rate for $\mu \rightarrow e \gamma$ directly to the sub-leading 
Yukawa couplings. We shall discuss the conditions under
which this may be possible later.
There are several examples of models in the literature which rely
on SD \cite{Altarelli:2002hx,Barbieri:1999pe,King:2000ge,King:2001uz}.
In the realistic models the dominant right-handed neutrino is either 
the heaviest or the lightest. We refer these cases as
heavy sequential dominance (HSD) or light sequential dominance (LSD),
respectively, and  discuss these cases separately.

In this paper, then, we perform a global analysis
of HSD and LSD models in the framework of the CMSSM,
presenting our predictions in the $(m_0,M_{1/2})$ plane, assuming
radiative electroweak symmetry breaking, and including
all observed laboratory data. We give predictions for branching fractions for 
$\tau \rightarrow \mu \gamma$ and $\mu \rightarrow e \gamma$ in terms
of a convenient parametrisation which we introduce for the
neutrino Yukawa couplings.
Our results confirm that the $\tau \rightarrow \mu \gamma$
rate distinguishes HSD from LSD. We further show 
that the $\mu \rightarrow e \gamma$ rate
may allow a determination of the sub-dominant
neutrino Yukawa couplings in the flavour basis,
in terms of an expansion parameter $\lambda$.
We show that
measurement of $\theta_{13}$ may determine a ratio of sub-dominant
Yukawa couplings. We discuss
quantitative effects which were not present in the 
leading log and leading mass insertion approximations of 
\cite{Lavignac:2001vp}, for example cases where
$\mu \rightarrow e \gamma$ is controlled by the 13 slepton mass.

The remainder of the paper is set out as follows.
In section 2 we discuss sequential dominance and introduce our
parametrisation. We also make some leading log analytic estimates of 
off-diagonal slepton masses in terms of this parametrisation,
and discuss the implications for LFV processes.
In section 3 we give our full numerical results.

\section{Sequential Dominance}

\subsection{Brief Review}

In this subsection we give a brief review of SD, including the
analytic estimates of neutrino masses and mixing angles in terms
of Yukawa couplings and right-handed neutrino masses, which 
we assume here to be real.
More details can be found in \cite{King:2002qh}.
Note that all the results in this subsection are independent of the 
mass ordering of right-handed neutrino masses, and so apply to both
HSD and LSD.

In the flavour basis where the charged lepton mass matrix is diagonal,
the diagonal right-handed neutrino mass matrix is written as
\begin{equation}
M_{RR}=
\left( \begin{array}{ccc}
X' & 0 & 0    \\
0 & X & 0 \\
0 & 0 & Y
\end{array}
\right) 
\label{srhnd}
\end{equation}
and the neutrino Yukawa matrix without loss of generality as
\begin{equation}
Y^{\nu}_{LR}=
\left( \begin{array}{ccc}
a' & a & d    \\
b' & b & e \\
c' & c & f
\end{array}
\right). 
\label{dirac}
\end{equation}
In order to account for a neutrino mass hierarchy and large neutrino
mixing angles in a natural way (without fine-tuning) the following
SD condition was proposed \cite{King:1999mb},
\beq
\frac{|e^2|,|f^2|,|ef|}{Y}\gg
\frac{|xy|}{X} \gg
\frac{|x'y'|}{X'}
\label{seq}
\eeq
where $x,y\in \{a,b,c\}$ and $x',y'\in \{a',b',c'\}$.
It is further assumed \cite{King:1998jw} that 
\beq
d\ll e\approx f.
\label{def}
\eeq
Then it was shown that the neutrino masses
are given to leading order in $m_2/m_3$
by \cite{King:2002qh},
\bea
m_1 & \sim & O(\frac{x'y'}{X'}\,v_2^2) \nonumber \\
m_2 & \approx & \frac{a^2+(c_{23}b -s_{23}c)^2}{X}\:v_2^2 \nonumber \\
m_3 & \approx & \frac{e^2+f^2}{Y}\:v_2^2
\label{masses}
\eea
where $v_2$ is a Higgs vacuum expectation value (vev) associated with
the (second) Higgs doublet that couples to the neutrinos.
Note that with SD each neutrino mass is generated
by a separate right-handed neutrino, and the origin of the neutrino mass
hierarchy is thus linked to the sequential condition in Eq.\ref{seq}.
The neutrino mixing angles are given to leading order in $m_2/m_3$
by \cite{King:2002qh},
\bea
\tan \theta_{23} & \approx & \frac{e}{f}\nonumber \\
\tan \theta_{12} & \approx &
\frac{a}{(c_{23}b -s_{23}c)} \nonumber \\
\theta_{13} & \approx & 
\frac{a(s_{23}b+c_{23}c)}{(e^2+f^2)}\frac{Y}{X} 
\label{angles}
\eea
Thus, assuming SD, the atmospheric angle is given 
by a ratio of dominant right-handed neutrino couplings associated with $Y$,
the solar angle is given by a ratio of sub-dominant right-handed neutrino
couplings associated with $X$, and $\theta_{13}$ is of order $m_2/m_3$.
The sub-sub-dominant right-handed neutrino $X'$ with couplings
$a',b',c'$ is completely irrelevant for neutrino physics
unless $m_1$ can be measured.
In writing the equation for $\theta_{13}$ we have assumed
for simplicity that 
\beq
d\ll {a(s_{23}b+c_{23}c)}Y/({\sqrt{e^2+f^2}}X).
\eeq
%For simplicity d\ll {a(s_{23}b+c_{23}c)}Y/({\sqrt{e^2+f^2}}X)$
%was assumed when writing the equation for $\theta_{13}$.
%The more general result including $d$ is given in \cite{King:1999mb}.
%In the opposite limit that 
%$d\gg {a(s_{23}b+c_{23}c)}Y/({\sqrt{e^2+f^2}}X)$,
%but still maintaining the condition $d\ll e\approx f$, 
%$\theta_{13}$ is described by the simpler result \cite{King:1998jw},
%\beq
%\theta_{13}\approx \frac{d}{\sqrt{e^2+f^2}}.
%\label{larged}
%\eeq

At leading order in a mass insertion approximation 
\cite{Borzumati:1986qx,Hisano:1995cp}
the branching fractions of LFV processes are given by
\beq
{\rm BR}(l_i \rightarrow l_j \gamma)\approx 
        \frac{\alpha^3}{G_F^2}
        f(M_2,\mu,m_{\tilde{\nu}}) 
        |m_{\tilde{L}_{ij}}^2|^2 \xi_{ij}  \tan ^2 \beta
    \label{eq:BR(li_to_lj)}
\eeq
where $l_1=e, l_2=\mu , l_3=\tau$,
and where the off-diagonal slepton doublet mass squared is given 
in the leading log approximation (LLA) by
\beq
m_{\tilde{L}_{ij}}^{2(LLA)}
\approx -\frac{(3m_0^2+A_0^2)}{8\pi ^2}C_{ij}
\label{lla}
\eeq
where the leading log coefficients are given by
\bea
C_{21} & = & a'b'\ln \frac{M_U}{X'} + ab\ln \frac{M_U}{X} 
+de\ln \frac{M_U}{Y} \nonumber \\
C_{32} & = & b'c'\ln \frac{M_U}{X'} + bc\ln \frac{M_U}{X} 
+ef\ln \frac{M_U}{Y} \nonumber \\
C_{31} & = & a'c'\ln \frac{M_U}{X'} + ac\ln \frac{M_U}{X} 
+df\ln \frac{M_U}{Y} 
\label{Cij}
\eea
The factors $\xi_{ij}$ in Eq.\ref{eq:BR(li_to_lj)}
represent the ratio of the leptonic partial
width to the total width, 
\beq
\xi_{ij}=\frac{\Gamma (l_i\rightarrow l_j\nu_i \overline{\nu}_j)}
{\Gamma (l_i \rightarrow {\rm all})}
\eeq
Clearly $\xi_{21}=1$ but $\xi_{32}$ is non-zero and must be included
for correct comparison with the experimental limit on the branching
ratio for $\tau \rightarrow \mu \gamma$. This factor is frequently forgotten
in the theoretical literature.

We shall focus on $C_{21}$ and $C_{32}$ which correspond to 
$\mu \rightarrow e\gamma$ and $\tau \rightarrow \mu \gamma$.
For HSD the couplings $a',b',c'$ are expected to be smaller
than the couplings $a,b,c$, and we will therefore
drop them in both our analytic and numerical analysis.
For LSD the primed terms are also not relevant for LFV since
as discussed later we will
set $X'=M_{U}$ which means that this right-handed neutrino is
immediately decoupled at the GUT scale, and so $\ln \frac{M_U}{X'}=0$.
In the case of LSD this is rather a strong assumption and 
if $X'<M_{U}$ these Yukawa couplings will again become relevant.
Dropping the primed contributions for both HSD and LSD the
relevant coefficients are then given by
\bea
C_{21} & = & ab\ln \frac{M_U}{X} +de\ln \frac{M_U}{Y} \nonumber \\
C_{32} & = & bc\ln \frac{M_U}{X} +ef\ln \frac{M_U}{Y}
\label{C2131}
\eea

Note that the primed couplings are irrelevant to either 
the neutrino masses and mixing angles or the LFV
processes. This applies to both the HSD and the LSD cases.

\subsection{A convenient parametrisation}

The above analytic results for the neutrino masses and mixing angles
suggest the following parametrisation of the Yukawa couplings and
right-handed neutrino masses in terms of order unity coefficients 
$a_{ij}$ and $A$, together with an expansion parameter
$\lambda$ raised to integer powers,
\bea
a & \equiv & a_{12}\lambda^nf \nonumber \\
b & \equiv & a_{22}\lambda^nf \nonumber \\
c & \equiv & a_{32}\lambda^nf \nonumber \\
d & \equiv & a_{13}\lambda^mf \nonumber \\
e & \equiv & a_{23}f \nonumber \\
X & \equiv & A\lambda^{2n-1}Y \nonumber \\
\label{param}
\eea
where 
\beq
\lambda \equiv \sqrt{\frac{|\Delta m_{21}^2|}{|\Delta m_{32}^2|}}\approx 0.15.
\label{lambda}
\eeq
Currently the data prefers $\lambda\approx 0.15$, with large uncertainty.
It is the smallness of this value that is the motivation for sequential
dominance. 

In terms of the parametrisation in Eq.\ref{param}, the
neutrino mixing angles
in Eq.\ref{angles} become, 
\bea
\tan \theta_{23} & \approx & a_{23}\nonumber \\
\tan \theta_{12} & \approx &
\frac{a_{12}}{a_{22}(c_{23} -s_{23}r)} \nonumber \\
\theta_{13} & \approx & 
\frac{\lambda}{A} \frac{a_{12}a_{22}(s_{23}+c_{23}r)}{(1+a_{23}^2)}
\label{angles1}
\eea
where for convenience we have also defined the following 
ratio of sub-dominant Yukawa couplings which clearly plays a crucial role
in determining $\theta_{13}$,
\beq
r=\frac{c}{b}=\frac{a_{32}}{a_{22}}.
\eeq
The motivation for the above parametrisation is that it reproduces 
the large atmospheric and solar mixing angles 
$\tan \theta_{23}\sim 1$, $\tan \theta_{12}\sim 1$, with the 
dimensionless parameters $a_{ij}$ being of order unity.
\footnote{
           Note that with all $a_{ij}$ and $A$ of order unity 
           this is not the most general parametrisation
that satisfies the conditions in Eqs.\ref{seq},\ref{def} that define
this class of models.
Also note that as before we have assumed for simplicity that 
$d$ is small in the equation for $\theta_{13}$, and the condition
for this is now $m\geq 2$.}

Similarly the neutrino masses in Eq.\ref{masses} become, 
in terms of the parametrisation in Eq.\ref{param},
\bea
m_1 & \approx & 0 \nonumber \\
m_2 & \approx & (a_{12}^2+a_{22}^2(c_{23} -s_{23}r)^2)
\frac{\lambda f^2}{AY}\nonumber \\
m_3 & \approx & (1+a_{23}^2)\frac{f^2}{Y} \label{masses1} 
\eea
The ratio of second and third neutrino masses is then
$m_2/m_3\sim \lambda$ which motivates the definition of the
expansion parameter in Eq.\ref{lambda}.
A further motivation for this parametrisation is that it clearly shows that
the neutrino masses and mixing angles in Eq.\ref{masses1}, \ref{angles1}
are completely independent of $n$.
Although the neutrino masses and mixings cannot determine 
the choice of integer $n$, which 
controls the magnitude of the Yukawa couplings, 
the LFV processes are able to do so as we now discuss.

Using our parametrisations in Eq.\ref{param} the coefficients relevant
for $\mu \rightarrow e\gamma$ and $\tau \rightarrow \mu \gamma$
from Eq.\ref{C2131} are then determined by the products of Yukawa couplings
\bea
ab & = & a_{12}a_{22}\lambda^{2n}f^2 \nonumber \\
de & = & a_{13}a_{23}\lambda^{m}f^2 \nonumber \\
bc & = & a_{22}^2r\lambda^{2n}f^2 \nonumber \\
ef & = & a_{23}f^2 
\label{ab}
\eea
Then from Eqs.\ref{C2131}, \ref{ab} we have
\bea
C_{21} & = & a_{12}a_{22}\lambda^{2n}f^2\ln \frac{M_U}{X} 
+a_{13}a_{23}\lambda^{m}f^2\ln \frac{M_U}{Y} \nonumber \\
C_{32} & = & a_{22}^2r\lambda^{2n}f^2 \ln \frac{M_U}{X} 
+a_{23}f^2 \ln \frac{M_U}{Y}
\label{C}
\eea
The coefficients $C_{ij}$ which determine the strength of LFV processes
are clearly governed by the value of the Yukawa coupling $f$ 
together with the integers $n,m$
which control the magnitudes of the Yukawa couplings.
The values of the right-handed neutrino masses 
does not directly influence the coefficients $C_{ij}$ very much,
with only a mild logarithmic dependence. However the values of the
right-handed neutrino masses will have an important indirect 
effect on LFV processes via the values of the Yukawa couplings.
For example for a fixed $m_3$, $C_{32}\propto f^2 \propto Y$,
so the $\tau \rightarrow \mu \gamma$ rate increases as  
the dominant right-handed neutrino mass $Y$ becomes heavier 
\cite{Blazek:2001zm}.
We shall now discuss the two cases HSD and LSD separately, corresponding to 
the dominant right-handed neutrino of mass $Y$ being the heaviest or the
lightest, respectively.

\subsection{Heavy Sequential Dominance (HSD)}

HSD is defined by the condition that the dominant right-handed
neutrino of mass $Y$ is the heaviest, with
\beq
Y\gg X \gg X'
\eeq
Typically in unified models the heaviest right-handed neutrino is in
the 33 position and the 33 element
of the neutrino Yukawa matrix is equal to the top quark 
Yukawa coupling $f=h_t$ (at the high energy scale). 
The HSD case then leads to the ``lop-sided'' form of neutrino Yukawa
matrix with an order unity Yukawa coupling in the 23 position,
and subsequently a large expected rate for
$\tau \rightarrow \mu \gamma$ \cite{Blazek:2001zm}. 
For HSD with $f=h_t$, the atmospheric neutrino mass 
$m_3\approx \sqrt{|\Delta m_{32}^2|}$ then implies that 
the heaviest (dominant) right-handed neutrino has mass 
\beq
Y\approx 2m_t^2/m_3\approx 3\times 10^{14} GeV. 
\eeq
where we have used the GUT scale value of the top quark mass $m_t$
which for large $\tan \beta$ is about 0.7 times its low energy value.
For HSD the condition that $X\ll Y$ implies, from Eq.\ref{param}, that
\beq
n\geq 1 
\label{n}
\eeq
Recall that we also assumed $m\geq 2$.
The neutrino matrices in Eqs.\ref{srhnd},\ref{dirac},
in terms of the parametrisation in Eq.\ref{param}, are summarised below
for the case of HSD,
\bea
M^{HSD}_{RR}&=&
\left( \begin{array}{ccc}
- & 0 & 0    \\
0 & A\lambda^{2n-1} & 0 \\
0 & 0 & 1
\end{array}
\right) Y \label{HSD1} \\
Y^{\nu HSD}_{LR}&=&
\left( \begin{array}{ccc}
- & a_{12}\lambda^n & a_{13}\lambda^m    \\
- & a_{22}\lambda^n & a_{23} \\
- & ra_{22}\lambda^n & 1
\end{array}
\right)h_t
\label{HSD2}
\eea
where the blanks indicate entries which are irrelevant
for both neutrino masses and mixing angles and LFV.
The neutrino masses and mixing angles are given by
Eqs.\ref{angles1},\ref{masses1}, independently of $n$.

In this case it is clear from Eq.\ref{C} that 
$\tau \rightarrow \mu \gamma$, which corresponds to $C_{32}$,
is large and independent of $n$, being given approximately by 
\beq
C_{32} = a_{22}h_t^2 \ln \frac{M_U}{Y}
\label{HSDC32}
\eeq
On the other hand $\mu \rightarrow e\gamma$, which corresponds to $C_{21}$,
is smaller and more model dependent since it depends on the
integers $n,m$,
\beq
C_{21}  =  a_{12}a_{22}\lambda^{2n}h_t^2\ln \frac{M_U}{X} 
+a_{13}a_{23}\lambda^{m}h_t^2\ln \frac{M_U}{Y} 
\label{HSDC21}
\eeq
From our numerical results we shall find that the experimental
limit on $\mu \rightarrow e\gamma$ will require $n\geq 2$.
Since $ab \sim \lambda^{2n}$ while $de \sim \lambda^{m}$,
if $2n<m$ then $ab$ will dominate over $de$, while if $2n>m$ then
$de$ will dominate over $ab$. In the second case
$\mu \rightarrow e\gamma$ is controlled by the 13 neutrino
Yukawa coupling $d$.
We emphasise the fact that although $\mu \rightarrow e\gamma$ is model
dependent in this framework we understand precisely its origin.

\subsection{Light Sequential Dominance (LSD)}

LSD is defined by the condition that the dominant right-handed
neutrino of mass $Y$ is the lightest, with
\beq
X'\gg X \gg Y
\eeq
The LSD case then leads a more 
``symmetrical'' form of neutrino Yukawa
matrix with no large off-diagonal Yukawa couplings, and hence a small
expected rate for $\tau \rightarrow \mu \gamma$.
Such a form of neutrino Yukawa matrix is consistent with 
an exactly symmetrical or anti-symmetrical or mixed symmetry form, 
or some more general matrix with small off-diagonal couplings. 
LSD is more complicated to discuss since we need to 
re-order the matrices in Eqs.\ref{srhnd},\ref{dirac},
to put the heaviest right-handed neutrino of mass $X'$ in the third column.
Then we shall make a similar assumption for LSD that the 33 element
of the neutrino mass matrix to be equal to the top quark 
Yukawa coupling $c'=h_t$.
For LSD to set the scale of the right-handed neutrino masses,
we need to relate the dominant right-handed neutrino Yukawa coupling
$f$ to $c'=h_t$. In particular, we extend our parametrisation to include
\beq
f=a_{31}\lambda^pc'
\label{f}
\eeq
where $p>0$, and $a_{31}$ is an order unity coefficient.
For LSD with $c'=h_t$, the atmospheric neutrino mass 
$m_3\approx \sqrt{|\Delta m_{32}^2|}$ then implies that 
the lightest (dominant) right-handed neutrino has mass 
\beq
Y\approx a_{31}^2\lambda^{2p}3\times 10^{14} GeV.
\eeq 
with $p>0$.
The sub-dominant right-handed neutrino mass is heavier than $Y$ and
its mass is parametrised by
\beq
X=A\lambda^{2n-1}Y\approx Aa_{31}^2\lambda^{2p+2n-1}3\times 10^{14} GeV
\label{X}
\eeq
with, 
\beq
n\leq 0<p. 
\label{np}
\eeq
In LSD the heaviest right-handed
neutrino mass is $X'$ which must be heavy enough to satisfy the 
sequential condition, corresponding to the second inequality in
Eq.\ref{seq},
\beq
\frac{|xy|}{X} \gg \frac{c'^2}{X'}
\label{seq2}
\eeq
From Eqs.\ref{param},\ref{X},\ref{seq2} we find
\beq
X'\gg \lambda^{-1}\, 3\times 10^{14} GeV
\label{X'}
\eeq
independently of $p,n$. 
Eq.\ref{X'} implies that $X'$ cannot be (much) below the GUT scale $M_{U}$.
As indicated earlier we shall assume $X'=M_{U}$ which implies that
there can be no LFV effects arising from this right-handed neutrino.
In order to ensure that $X\ll X'$ by comparing Eqs.\ref{X} and \ref{X'}
we also have the condition
\beq
p+n>0
\label{p+n}
\eeq

The neutrino matrices in Eqs.\ref{srhnd},\ref{dirac},
in terms of the parametrisation in Eqs.\ref{param},\ref{f} are summarised below
for the case of LSD, after re-ordering them to put the heaviest
right-handed neutrino in the third column, and the lightest (dominant)
right-handed neutrino in the first column,
\bea
M^{LSD}_{RR}&=&
\left( \begin{array}{ccc}
a_{31}^2\lambda^{2p} & 0 & 0    \\
0 & Aa_{31}^2\lambda^{2p+2n-1} & 0 \\
0 & 0 & -
\end{array}
\right) 3\times 10^{14} GeV \label{LSD1} \\
Y^{\nu LSD}_{LR}&=&
\left( \begin{array}{ccc}
a_{13}a_{31}\lambda^{p+m} & a_{12}a_{31}\lambda^{p+n} & -    \\
a_{23}a_{31}\lambda^{p} & a_{22}a_{31}\lambda^{p+n} & - \\
a_{31}\lambda^{p} & ra_{22}a_{31}\lambda^{p+n} & 1
\end{array}
\right)h_t
\label{LSD2}
\eea
where the blanks indicate entries which are irrelevant
if $X'=M_{U}$. However, as we mentioned previously, if $X'<M_{U}$ such Yukawa 
couplings will again become important.
Note that $n$ is a non-positive integer, and in realistic models
it is nearly always negative so that the Yukawa couplings in the
first column are smaller than the Yukawas in the second column.
The advantage of this parametrisation is that it provides
a unified parametrisation of HSD and LSD, with $n$ being positive
for HSD, and negative or zero for LSD. In this unified parametrisation
the neutrino masses and mixing angles are still given by the
same formulas as in Eqs.\ref{angles1},\ref{masses1},
idependently of $n,p,a_{31}$.

In this case it is clear from Eq.\ref{C} that 
$\tau \rightarrow \mu \gamma$, which corresponds to $C_{32}$,
is now much smaller and model dependent since it depends on the
integers $p,n$,
\beq
C_{32} =  ra_{22}^2a_{31}^2\lambda^{2p+2n}h_t^2\ln \frac{M_U}{X} 
+a_{22}a_{31}^2\lambda^{2p}h_t^2\ln \frac{M_U}{Y} 
\label{C322}
\eeq
If $n<0$ then, in the notation of Eq.\ref{ab},
$bc$ dominates over $ef$.
Turning to $\mu \rightarrow e\gamma$, which corresponds to $C_{21}$,
since $n\leq 0$ while $m\geq 2$ we have approximately,
\beq
C_{21}  =  a_{12}a_{22}a_{31}^2\lambda^{2p+2n}h_t^2\ln \frac{M_U}{X} 
\label{C212}
\eeq
so that, in the notation of Eq.\ref{ab}, $ab$ always dominates over $de$.
By comparing Eqs.\ref{C322} to \ref{C212}, we see that in all 
cases 
\beq
C_{32}\sim C_{21}\sim \lambda^{2p+2n}h_t^2\ln \frac{M_U}{X}
\label{prediction}
\eeq
which leads to the LSD prediction that the branching fractions for  
$\tau \rightarrow \mu \gamma$ and $\mu \rightarrow e\gamma$
should be comparable, and are controlled by the value of $p+n$.
We already saw in Eq.\ref{p+n} that $X\ll X'$ requires $p+n>0$.
We shall see in the next section that the experimental limit
on $\mu \rightarrow e\gamma$ will require $p+n\geq 2$.
This implies that the $\tau \rightarrow \mu \gamma$ rate for 
LSD is suppressed relative to that for HSD by of order $\lambda^8$
or more, which effectively makes $\tau \rightarrow \mu \gamma$ 
unobservable in this case. In the case that $X'<M_{U}$
and the 23 Yukawa coupling were set equal to the 32 Yukawa coupling
we would find 
\beq
C_{32}\sim \lambda^{p+n}h_t^2\ln \frac{M_U}{X'}
\label{prediction2}
\eeq
which would imply that the $\tau \rightarrow \mu \gamma$ rate for 
LSD is suppressed relative to that for HSD by of order $\lambda^4$
or more, which is still at least three orders of magnitude below current
limits.

\section{Numerical Results}
Our numerical results are based on the top-down global 
analysis of the CMSSM with universal soft scalar mass
$m_0$, soft gaugino mass $M_{1/2}$, soft trilinear mass $A_0$
and the sign of the Higgs superpotential mass parameter
$\mu$ which we take to be 
positive as implied by the recent muon anomalous magnetic moment
measurement. In a standard notation \cite{Blazek:1997}
the complete list of the input parameters varied at the 
unification scale $M_U\approx 2\times 10^{16}\,$GeV (which itself
is allowed to vary) contains the soft masses $m_0$, $M_{1/2}$, $A_0$, the
unified gauge coupling $\alpha_U$, the deviation of the QCD gauge
coupling $\alpha_3$ from $\alpha_U$,
$\epsilon_3\equiv (\alpha_3-\alpha_U)/\alpha_U$, the deviation 
of the $b,\tau$ Yukawa couplings $h_{b,\tau}$ from the (varying)
top Yukawa coupling $h_t$, $\epsilon_{b,\tau}=(h_{b,\tau}-h_t)/h_t$, 
and the five
parameters of the neutrino sector $Y$, $a_{12}$, $a_{22}$, 
$a_{23}$ and $A$ defined earlier.
The five neutrino parameters are only allowed 
to vary within $20-30\%$ (see Eq.\ref{param}) but typically 
good fits are obtained 
for values in the smaller range $\sim 10\%$ which means that
$Y$ and the $a$'s stay close to $3\times 10^{14}$GeV
and $1$, respectively. $r=a_{32}/a_{22}$ is held fixed 
in the analysis.\footnote
                         {
                          We study cases with $r$ taking on
                          different values as explained below.
                         }
$\mu$ is an input parameter at the low scale which we take 
as the $M_Z$ scale. 
The function $\chi^2 = \sum (x_i^{CMSSM}-x_i^{exp})^2/\sigma_i^2$ 
is evaluated based on the following observables $x_i\,$: 
$\alpha$, $G_\mu$, $\alpha_3(M_Z)$,  
$M_t$, $m_b(m_b)$, $M_\tau$, $M_Z$, $M_W$, $\Delta\rho$,
$BR(b\to s\gamma)$, $a_\mu^{\rm NEW}$ and the neutrino
mass differences and mixing angles of Eqs.(\ref{neutrinodata1}) 
and (\ref{neutrinodata}). 
The low-scale value of 
the bilinear parameter $B\mu$ is directly related to ratio of Higgs
vacuum expectation values 
$\tan\!\beta$ and the latter is kept fixed in the analysis.
Each computed charged fermion mass includes the complete 
 1-loop SUSY threshold correction and
correct radiative electroweak symmetry breaking at one loop is required 
at all points by adopting tight $\sigma_{M_Z}$ and $\sigma_{M_W}$.
A direct search limit is applied to the mass of each unobserved particle
including $m_{h^0}>114\,$GeV. 
We shall present our results as contours in the $(m_0,M_{1/2})$ plane,
for fixed $\tan\!\beta$,
with $A_0$ and $|\mu|$ varying over the plane.
More details of the global analysis can be found in 
\cite{Blazek:1997} and  \cite{Blazek:2002}. %%%%Here we mention that
\footnote
{
         We note that in the presented top-down analysis the first and second 
         generation quark and charged lepton masses and CKM elements 
         are always fit very well in a separate minimisation with small 
         first and second generation yukawa couplings as
         input at $M_U$. The presence of the $3\times 3$ 
         yukawa matrices $Y^u$, $Y^d$ and diagonal $Y^e$ renders 
         the whole analysis more complete and consistent
         while, at the same time, the additional flavour structure 
         of $Y^u$ and $Y^d$ has
         no effect on the rest of the analysis.
}

The main constraint of the present study comes from the muon $g-2$. 
Based on \cite{amu_ICHEP02,amu_SM} we require the contribution from new
physics to fit
\beq
a_{\mu}^{\rm NEW} = a_{\mu}^{\rm (exp)} - a_{\mu}^{\rm SM}
                  = (34\pm 11)\times 10^{-10}.
\label{amu_exp}
\eeq
%  
%  
%  
%  The universal trilinear parameter
%  $A_0$ is varied across the $m_0-M_{1/2}$
%  plane in order to mimimise the global $\chi^2$ of the fit.
%  The details of the global analysis will be discussed in 
%  a future publication \cite{Blazek:2002}. The main features
%  we mention here are that the quark masses and mixing angles
%  and the charged lepton masses, 
%  with the inclusion of the complete 1-loop SUSY threshold corrections, 
%  are treated as inputs, and the value of $|\mu|$ is fixed by the
%  requirement of 
%  correct radiative electroweak symmetry which is required at all points.
%  We also fit all remaining measured observables including 
%  $b\rightarrow s \gamma$ and the muon anomalous magnetic moment,
%  and ensure that the higgs boson mass satisfies its mass bound,
%  as will also be discussed elsewhere \cite{Blazek:2002}.
%  
%  
The main focus of the analysis is neutrino masses and mixing angles 
which we fit to the central atmospheric and LMA MSW values. 
$\theta_{13}$ is predicted in terms of a ratio of Yukawa couplings
as discussed above and computed from the exact form of the MNS matrix
obtained at low energy.  We emphasise that the neutrino
parameter fit is within the framework
of SD, and in this
framework it becomes possible to make direct connections between
LFV processes and specific Yukawa couplings, subject to the
conditions discussed in sections 2.3 and 2.4.
This would not be
possible for example if the hierarchy resulted from the tuning
of parameters. It is also worth noting that we perform
a careful RG analysis, adjusting the RG evolution below each
right-handed neutrino mass threshold, and also
perform an exact one-loop calculation of lepton flavour violating processes.
Thus we have very accurate predictions for the $\tau\to \mu\gamma$ and
$\mu\to e\gamma$ experiments which is increasingly important due
to the experimental progress. 
We shall compare our results to 
the leading log and mass insertion approximations described in the
previous sections, and in some cases find large discrepancies.

\begin{figure}[p]
\epsfysize=3.5truein
\epsffile{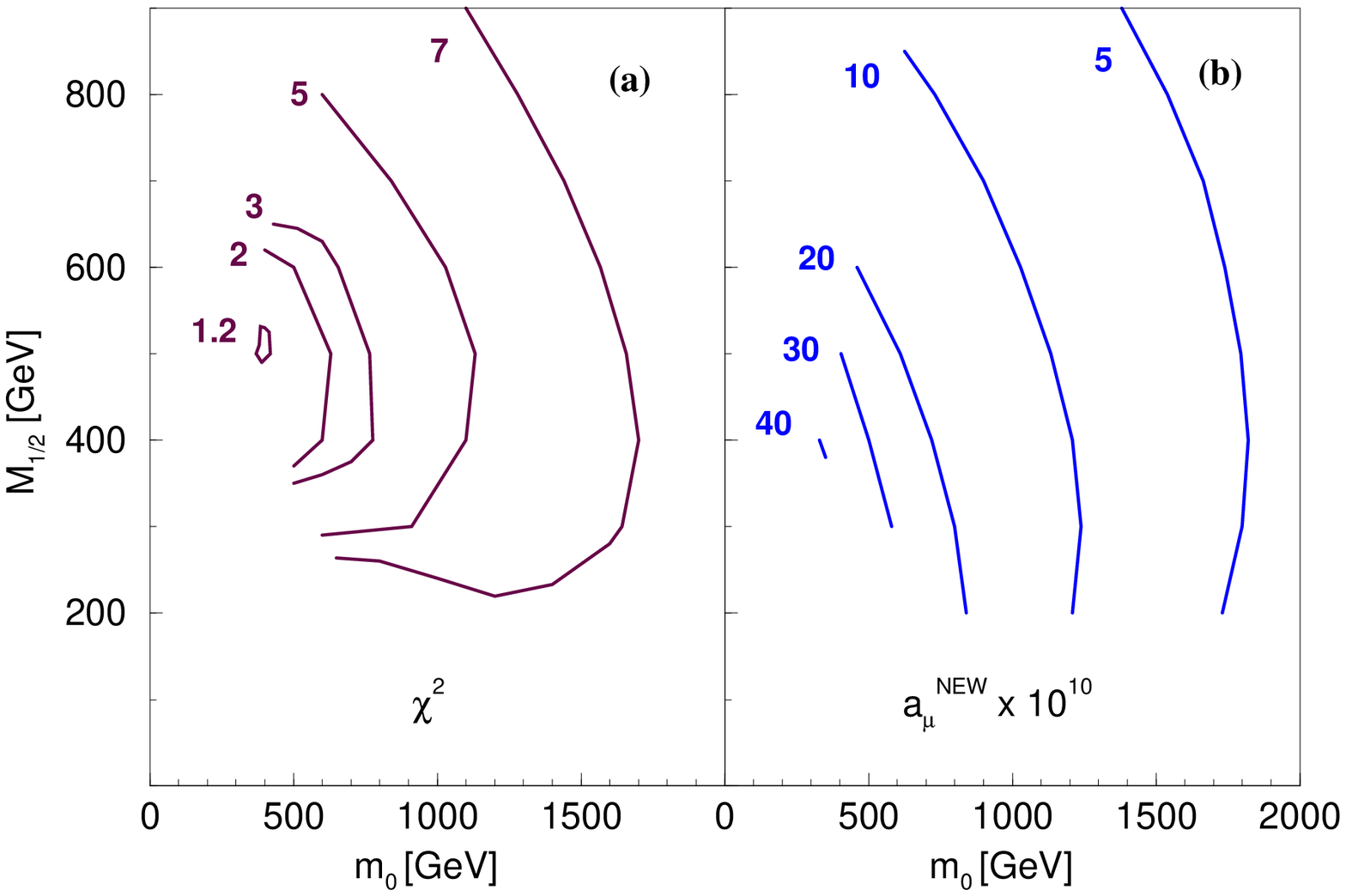}
\caption{Results for the global analysis of HSD.
We use the baseline parameters
$\tan \beta=50$, $n=2$, $r=-1$, $a_{13}=0$.
Panel (a) shows the $\chi^2$ of the fit which is minimised for 
$m_0\sim M_{1/2}\sim 500$ GeV. Panel (b) shows the SUSY
contribution to the anomalous magnetic moment of the muon.}
\label{baseline}
\end{figure}

In Figure \ref{baseline} we show the results of a global analysis
of the HSD model for $\tan \beta =50$, $n=2$, $r=-1$, $a_{13}=0$,
and the parameters refer to the matrices in Eq.\ref{HSD1},\ref{HSD2}. 
%  The remaining parameters
%  $A,a_{23},a_{12},a_{22}$ are allowed to vary over the range
%  $0.7-1.3$ (typically good fits are obtained for values 
%  in the smaller range $0.9-1.1$). 
This choice of parameters we call the ``baseline'' parameter set. 
Figure \ref{baseline}(a) shows the global $\chi^2$ 
fit across the $(m_0,M_{1/2})$ plane, with the best fit in the region 
$m_0\sim M_{1/2}\sim 500$ GeV.
The fit deteriorates for large $M_{1/2}$ and $m_0$ because of the muon 
anomalous magnetic moment which is shown in the panel (b) 
and for small $M_{1/2}$ and small $m_0$ because of $b\to s\gamma$.
In the first case the sparticles are too heavy to generate the 
observed discrepancy (\ref{amu_exp}) 
while in the latter case they are too light
making the chargino contribution to the $b\to s \gamma$ effective
amplitude too large. 

\begin{figure}[p]
\epsfysize=6.5truein
\epsffile{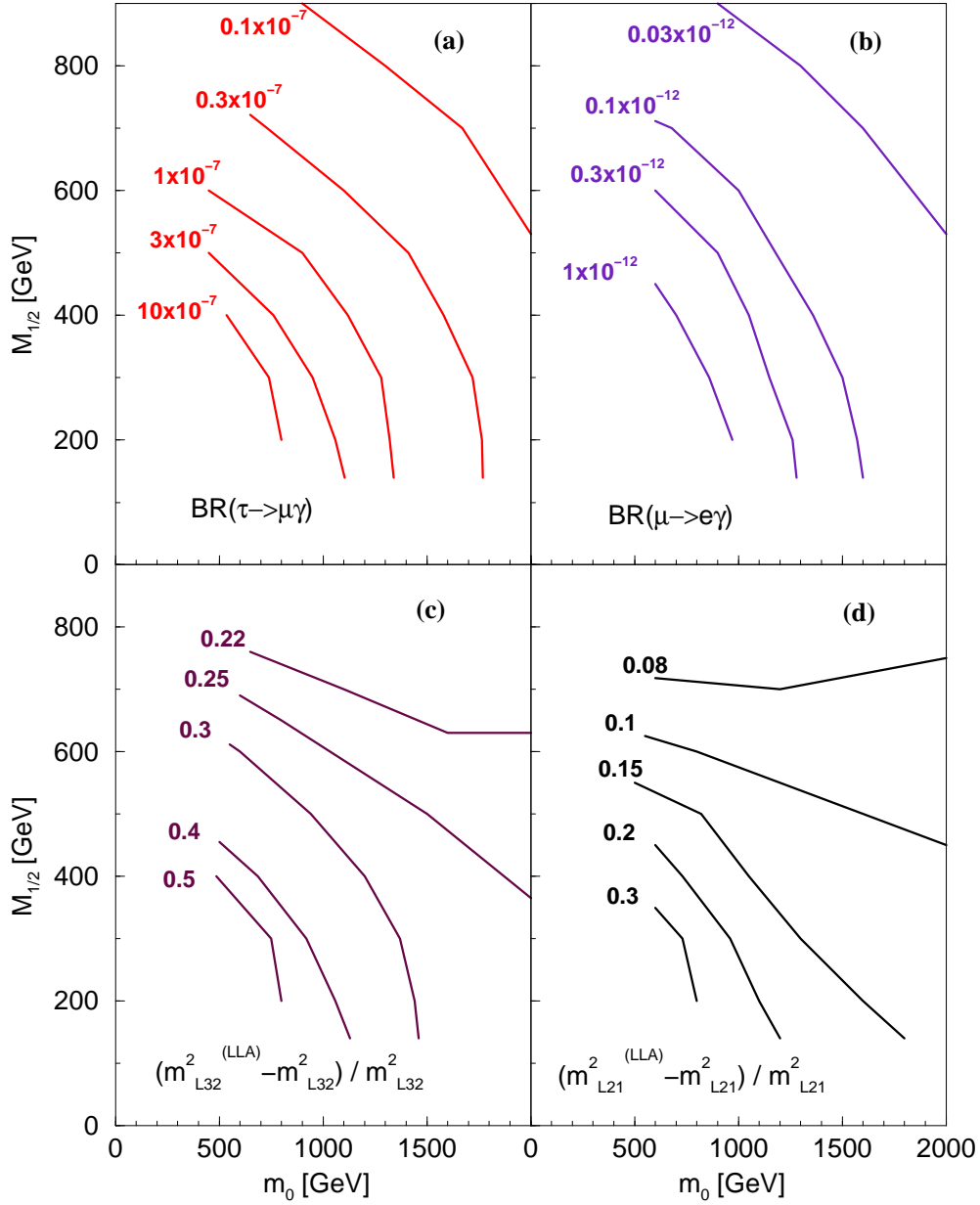}
\caption{The upper panels show the predictions for the branching fraction 
for (a) $\tau \rightarrow \mu \gamma$ and (b) $\mu \rightarrow e\gamma$
for HSD using the same baseline parameters as in 
Figure \ref{baseline}.
The lower panels (c), (d)
show the fractional error $\Delta_{ij}$, defined in the
text, that would be
made in calculating the off-diagonal slepton masses if the leading log
approximation had been used instead of the exact calculation.}
\label{LFVHSD}
\end{figure}

In Figure \ref{LFVHSD} we show the predictions of the HSD model with the
baseline parameters for the branching ratios of 
$\tau \rightarrow \mu \gamma$ and $\mu \rightarrow e\gamma$.
As expected from Eqs.\ref{HSDC32} and \ref{HSDC21} the upper panels
show a large rate for (a) $\tau \rightarrow \mu \gamma$ 
and (b) $\mu \rightarrow e\gamma$ close to the current limits. 
The prediction for the large $\tau \rightarrow \mu \gamma$ rate 
in panel (a) is quite a
robust prediction of HSD \cite{Blazek:2001zm}. As discussed later the
predictions are expected to scale as $\tan^2 \beta$.
The $\mu \rightarrow e\gamma$ rate in panel (b) is sensitive to
the choice of $n$, which we have taken to be $n=2$.
This rate can easily be enhanced by taking
smaller $n$ or suppressed by taking larger $n$, as is clear from
Eq.\ref{HSDC21}. In the lower panels of Figure \ref{LFVHSD} we investigate the
accuracy of the leading log approximation (LLA) used to calculate the
off-diagonal slepton masses in Eq.\ref{lla}, with the HSD coefficients
in Eqs.\ref{HSDC32},\ref{HSDC21}, as compared to the exact 
numerical calculation used to generate our numerical results.
We define the fractional error as follows:
\beq
\Delta_{ij}\equiv
\frac{m_{\tilde{L}_{ij}}^{2(LLA)}-m_{\tilde{L}_{ij}}^2}
{m_{\tilde{L}_{ij}}^2}
\eeq
Figure \ref{LFVHSD}(c) shows contours
of $\Delta_{32}$ %%%%corresponding 
while panel (d) shows
$\Delta_{21}$, in each case giving a measure of the error that would have
been incurred had the LLA been used in calculating the 
rates for $\tau \rightarrow \mu \gamma$ and $\mu \rightarrow e\gamma$.
The rates shown in the upper panels were, of course,  calculated exactly
without using the LLA approximation. The LLA can induce errors of up to 
50\% for the lighter SUSY masses. Note that the errors double for
the branching ratio. 

\begin{figure}[p]
\epsfysize=6.5truein
\epsffile{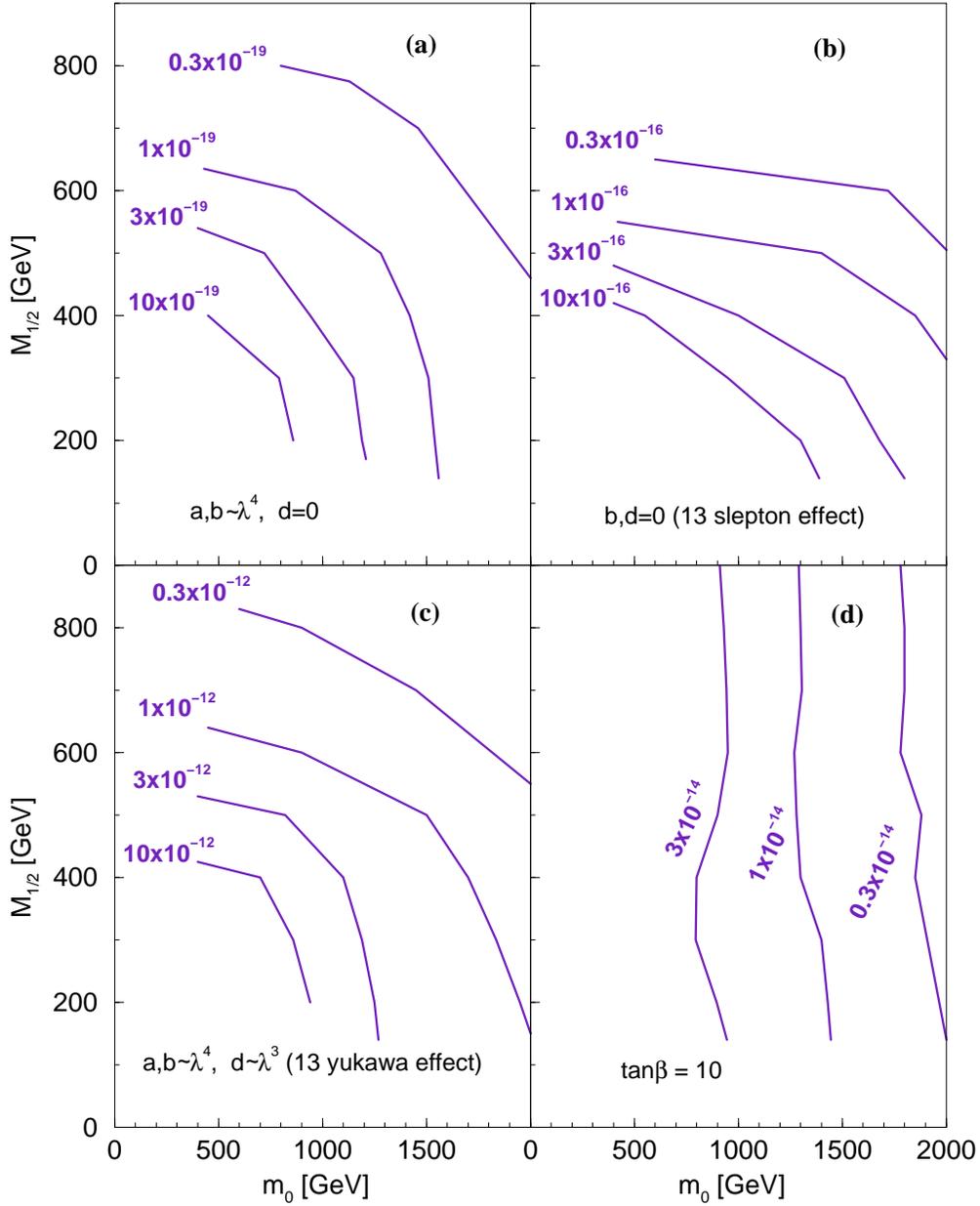}
\caption{Predictions for BR($\mu\to e\gamma$) in HSD for alternative
choices of parameters in this class of models. In panel (a),
we choose $\tan \beta=50$, $n=4$, $r=-1$, $a_{13}=0$.
In panel (b) we choose 
$\tan \beta=50$, $n=2$, $r=0$, $a_{13}=0$, $a_{22}=0$.
In panel (c) we choose 
$\tan \beta=50$, $n=4$, $r=-1$, $m=3$.
In panel (d) we choose
$\tan \beta=10$, $n=2$, $r=-1$, $a_{13}=0$.}
\label{muegamma}
\end{figure}

The parameter dependence of BR($\mu\to e\gamma$)
in the HSD model is examined in 
Figure \ref{muegamma}, where in each case we allow one or
two parameters to change from the baseline parameter set used
in Figures \ref{baseline} and \ref{LFVHSD}, 
in order to illustrate a particular effect. 
In Figure \ref{muegamma}(a),
we suppress the subdominant Yukawa couplings by taking $n=4$,
leaving all the other baseline parameters unchanged.
As expected from Eq.\ref{HSDC21}, this has the effect of suppressing
BR($\mu\to e\gamma$) by a relative factor of 
$\lambda^8\approx 2.5\times 10^{-7}$ compared to the previous result, 
which approximately corresponds to what is observed by comparing
Figure \ref{muegamma}(a) to 
Figure \ref{LFVHSD}(b).
This result demonstrates how sensitive BR($\mu\to e\gamma$)
is to the values of the subdominant
Yukawa couplings of order $\lambda^n$, with the prospect 
that a measurement of this rate is equivalent to a measurement
of these Yukawa couplings parametrised by $n$. This is an exciting
prospect offered by the SD class of models, however this result
is subject to the discussion below Eq.\ref{HSDC21}, which 
we shall return after the following paragraph.

We now discuss the numerical importance of effects
beyond the leading mass insertion approximation.
In Figure \ref{muegamma}(b) we set
$a_{22}=0$ with the other parameters as in the baseline set,
in particular $a_{13}=0$. Having $a_{22}=a_{13}=0$ should completely 
kill $\mu \rightarrow e\gamma$ according to Eq.\ref{HSDC21},
however clearly it does not, since the rate in 
Figure \ref{muegamma}(b) is in fact larger than the rate in
Figure \ref{muegamma}(a) by three orders of magnitude!
Clearly the mass insertion approximation, on which our analytic
expections of the previous section, and all the results in 
\cite{Lavignac:2001vp} for example, are based, is no longer reliable
in the limit of $a_{22}=a_{13}=0$.
We note that having $a_{22}=a_{13}=0$ is completely reasonable from
the SD point of view of neutrino masses and mixing angles, and would
correspond to some specific type of texture of the ``Fritzsch'' type
\cite{Fritzsch:1999ee} for example. 
Although in this case $C_{21}=0$, the values of $C_{32}$ and
$C_{31}$ are non-zero, leading to off-diagonal 32 and 31 slepton
masses, where according to Eq.\ref{Cij} the 31 slepton mass
must be arising from the $ac$ product of Yukawa couplings
which would be suppressed by increasing $n$.
A 21 flavour violation may be generated by a double mass
insertion involving both 32 and 31, and this results in the 
observed rate for $\mu \rightarrow e\gamma$ in
panel (b) of Figure \ref{muegamma}, although we calculate the
rate exactly without relying on any mass insertion methods
at all as in \cite{King:1998nv, Blazek:2001zm}, for example.
We call this the 13 slepton effect.

We now consider the effects of the 13 Yukawa coupling which
contribute to BR($\mu\to e\gamma$) according to 
the second term of Eq.\ref{HSDC21}.
Such effects are illustrated in
Figure \ref{muegamma}(c) where we keep $n=4$ as in panel (a),
but now take $a_{13}$ to be of order one, with the 13 Yukawa coupling
controlled by the value of $m$, which we take to be $m=3$.
Switching on the 13 Yukawa coupling in Figure \ref{muegamma}(c) 
results in a dramatic increase in BR($\mu\to e\gamma$)
compared to Figure \ref{muegamma}(a). We call this the 13 Yukawa effect.

Figure \ref{muegamma}(d) shows BR($\mu\to e\gamma$)
in the HSD model where $\tan \beta =10$
and the remaining parameters take the baseline values. 
Comparing to the baseline contours in Figure \ref{baseline}(b), 
it is clear that the contours do not simply scale as $\tan^2 \beta$.
The reason is that $|\mu|$ is fixed by electroweak symmetry breaking,
and also depends on $\tan \beta$, and this leads to the difference
in the two sets of contours. If $|\mu|$ were held fixed then we would
find the expected $\tan^2 \beta$ scaling of the rates. The figures
illustrate the importance of the implicit $|\mu|$ dependence
for differing values of $\tan \beta$, and warn against naively
assuming a $\tan^2 \beta$ scaling of the results in the 
$(m_0,M_{1/2})$ plane.

\begin{figure}[p]
\epsfysize=6.5truein
\epsffile{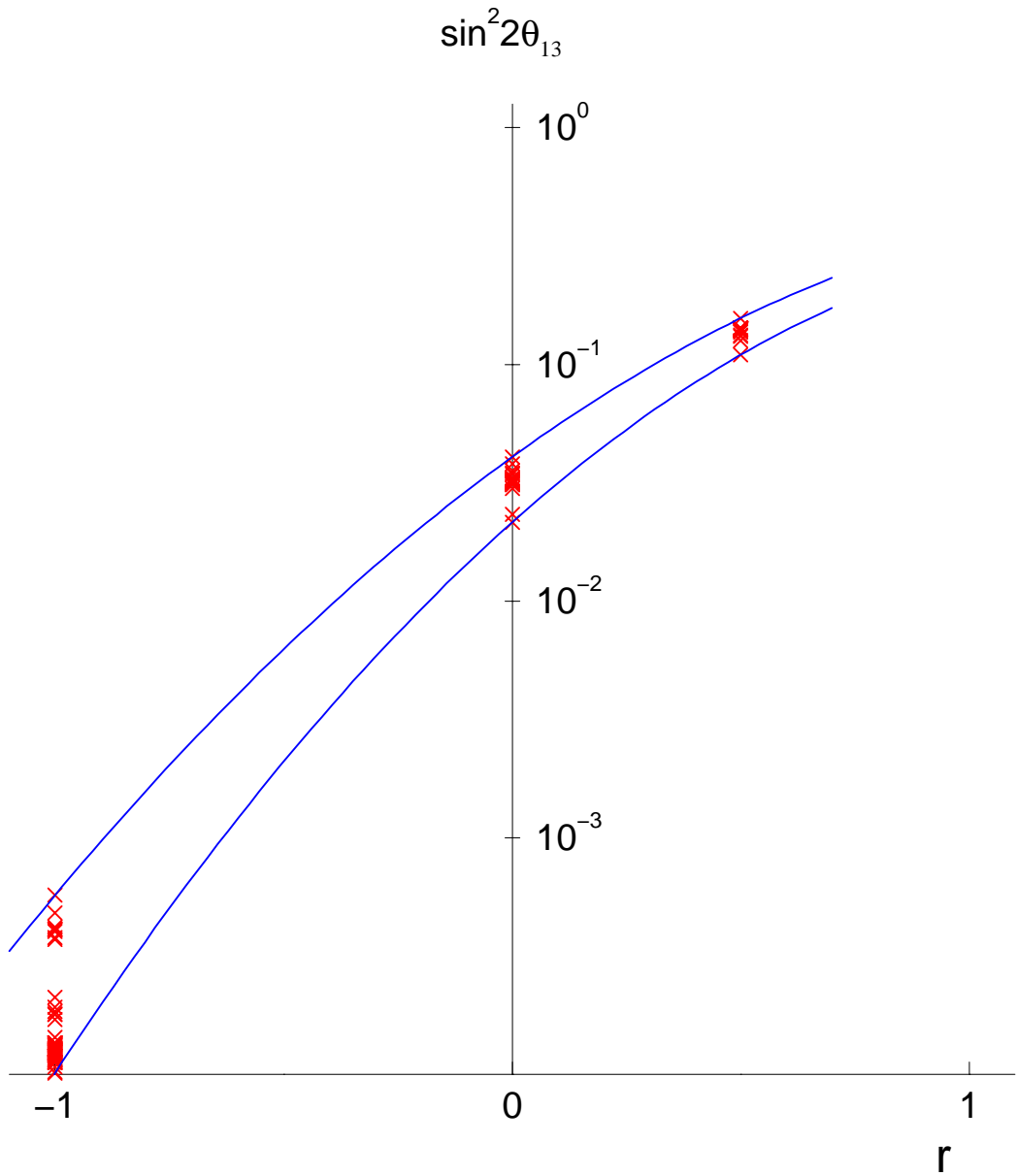}
\caption{The prediction of $\sin^22\theta_{13}$ as a function of 
the ratio of subdominant Yukawa couplings $r=a_{32}/a_{22}$
for the HSD class of models. The remaining parameters
are as in Figure \ref{baseline}.}
\label{theta}
\end{figure}

In Figure \ref{theta} we show 
the prediction of $\sin^22\theta_{13}$ as a function of 
the ratio of subdominant Yukawa couplings $r=a_{32}/a_{22}$
for the HSD class of models, taking all the other parameters
equal to their baseline values, in particular $a_{13}=0$.
The qualitative variation of this angle with $r$ follows from
Eq.\ref{angles1} which shows that although 
the natural value of the angle $\theta_{13}$ is set
by the parameter $\lambda$, as
$r\rightarrow -1$, $\theta_{13}\rightarrow 0$,
while maintaining $\tan \theta_{12}\sim 1$. 
Numerically we perform a 
global fit for three specific values of $r$
and for each value we find a band of values
of $\sin^22\theta_{13}$ corresponding to the variation in the
values of the order one coefficients.
For $r=-1$ the value of $\sin^22\theta_{13}$ is not exactly equal to 
zero, due to subleading effects beyond the level of approximation 
of Eq.\ref{angles1}, but it does get very small.
In general it is fair to say that over most of the range of $r$
the value of $\sin^22\theta_{13}$ is below the current CHOOZ limit
of about 0.14 but for $r>0$ it should be within the combined 
ICARUS and OPERA expected 2009 limit of 0.03. However it is
clear that for $r<0$ a measurement of $\sin^22\theta_{13}$ would
only be possible at JHF or a Neutrino Factory.
Note that Figure \ref{theta} assumes $a_{13}=0$, and the result
is more complicated if this is not the case.
We emphasise that for $a_{13}=0$
the rate for $\mu \rightarrow e\gamma$ is generally
independent of the prediction for $\theta_{13}$, which is clear
analytically by comparing Eq.\ref{angles1} to Eq.\ref{HSDC21}.
The point is simply that $C_{21}$ is independent of $a_{32}$ and 
hence $r$. 
\footnote
{
  In the finite 13 Yukawa case
  $\theta_{13}$ is given by a more complicated formula \cite{King:2002qh}.
  However, the experimental limit on $BR(\mu \rightarrow e\gamma)$
  is a strong constraint in this case. 
}
We emphasise that $\theta_{13}$ can be rather large 
while $\mu \rightarrow e\gamma$ can be arbitrarily suppressed
simply by increasing the value of $n$.

\begin{figure}[p]
\epsfysize=6.5truein
\epsffile{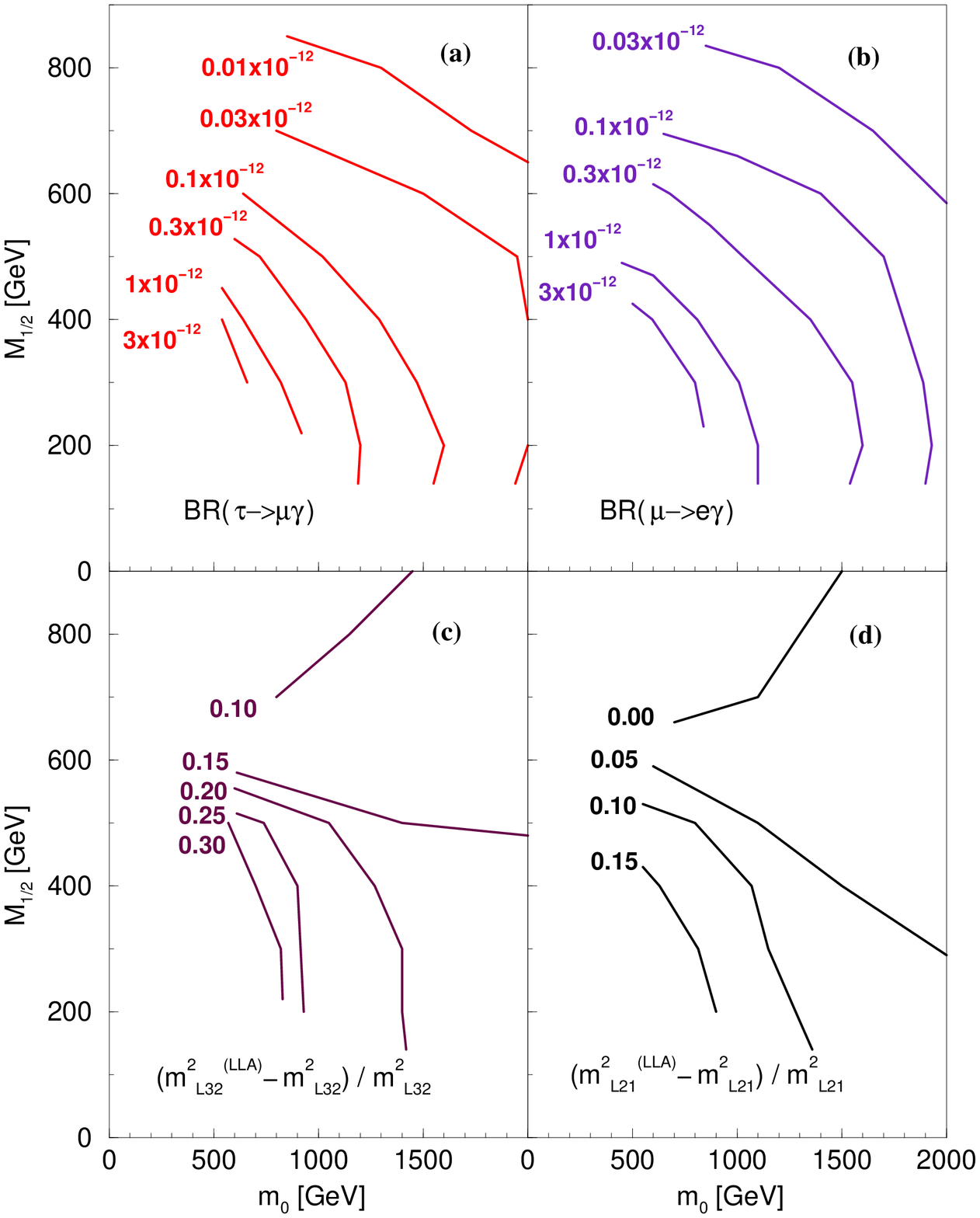}
\caption{The upper panels show the predictions for the branching fraction 
for (a) $\tau \rightarrow \mu \gamma$ and (b) $\mu \rightarrow e\gamma$
for the LSD class of models.
The parameters used are $\tan \beta=50$, $p+n=2$, $n=-1$, $r=-1$, $a_{13}=0$.
The lower panels (c), (d) show the fractional error $\Delta_{ij}$, defined
in the text, that would be
made in calculating the off-diagonal slepton masses if the leading log
approximation had been used instead of the exact calculation.}
\label{LFVLSD}
\end{figure}

In Figure \ref{LFVLSD} we show predictions of the LSD model 
for the branching ratios of 
$\tau \rightarrow \mu \gamma$ and $\mu \rightarrow e\gamma$.
The parameters used are $\tan \beta=50$, $p+n=2$, $n=-1$, $r=-1$, $a_{13}=0$,
where the parameters refer to Eqs.\ref{LSD1},\ref{LSD2}.
As before, the remaining parameters
$A,a_{ij}$ are allowed to vary over the range
$0.7-1.3$ (and again typically good fits are obtained for values 
in the smaller range $0.9-1.1$).
The LSD prediction for BR($\mu\to e\gamma$) in
Figure \ref{LFVLSD}(b) is comparable to that 
for the HSD model in Figure \ref{LFVHSD}(b).
This is expected since the coefficient $C_{21}$
for the LSD model in Eq.\ref{C212} with $p+n=2$ is of similar
magnitude to the coefficient $C_{21}$
for the HSD model in Eq.\ref{HSDC21} with $n=2$.
The new feature in the LSD model is that the
magnitude of BR($\tau\to \mu\gamma$) in Figure \ref{LFVLSD}(a) 
is now much smaller, and in fact is similar to 
the magnitude of BR($\mu\to e\gamma$) in Figure \ref{LFVLSD}(b), 
as predicted by Eq.\ref{prediction}.
However, if we had allowed $X'<M_{U}$ then a larger, but still
suppressed, prediction for $\tau \rightarrow \mu \gamma$ following from
Eq.\ref{prediction2} would have been found.
In any case the practical conclusion is the same: in the forseeable future
$\tau \rightarrow \mu \gamma$ is experimentally unobservable for the
LSD case. However $\mu \rightarrow e\gamma$ 
may be observable and is determined by 
the choice of $p+n$, which we have taken to be $p+n=2$.
Figures \ref{LFVLSD}(c),(d) show 
the accuracy of the LLA for the LSD class of models.
As before the accuracy improves
for larger SUSY masses but the error doubles in the evaluation
of the branching ratio. 

Note that in the case of the LSD class of models
the branching ratio for $\mu \rightarrow e\gamma$ 
can easily be enhanced by taking
smaller $p+n$ or suppressed by taking larger $p+n$, as is clear from
Eq.\ref{C212}. Thus, a measurement of this branching ratio gives
a measurement of the subleading Yukawa couplings of the second 
column, as in the HSD case. However, unlike the HSD case,
the LSD case is not sensitive to the 13 slepton effect and the 
13 Yukawa effect in Figures \ref{muegamma}(b),(c). 
The effects beyond the leading mass
insertion, due to a double slepton mass insertion, and 
effects coming from the Yukawa coupling parametrised by $a_{31}$,
are both negligible for the LSD case. The reason that these
effects were important for the HSD case was due to the large,
order unity, Yukawa coupling in the 23 position. In the LSD
case there are no large Yukawa couplings apart from that in the
33 position, and hence such effects are not so important in this case.
This means that in the LSD case, the measurement of 
BR($\mu\to e\gamma$) 
is more reliably related to a measurement of the Yukawa
couplings parametrised by $p+n$.

\section{Summary}

This paper has focused on models which satisfy the
sequential dominance condition. 
We have considered separately two broad classes of models,
HSD and LSD, in which the dominant right-handed neutrino
is the heaviest or lightest one. Although specific examples
of models exist in the literature which satisfy either the HSD 
\cite{King:2000ge} or LSD \cite{King:2001uz} conditions, 
we emphasise that the sequential dominance condition
is simply equivalent to assuming that a neutrino mass hierarchy
is generated via the see-saw mechanism without
fine-tuning in the leptonic sector,
in the presence of large atmospheric and solar mixing angles.
Therefore our results in fact apply to a very large class of models,
although some of our specific results apply only in certain
specific regions of parameter space.

We have presented approximate
analytic results for neutrino masses and mixing angles,
and for off-diagonal slepton masses in the leading log
approximation in each case, using a new parametrisation
of the matrices at the high energy scale. The parametrisation
is very useful in relating the neutrino and LFV observables
and in interpreting the data in terms of model parameters.

The numerical results, extracted from the best fits of the
global top-down analysis, demonstrate the validity (limited,
at times) of the leading log and leading mass insertion
approximations used in the literature. We find that the magnitude of
BR($\tau\to\mu\gamma$) provides the main discriminator between 
the different classes of sequential dominance, with a large
rate not far below current limits predicted by HSD, and a much smaller
rate predicted by LSD. The observation of 
BR($\mu\to e\gamma$), on the other hand, may determine the order
of the sub-dominant neutrino Yukawa couplings in the flavour
basis in both the HSD and LSD cases. However, due to the large
23 Yukawa coupling, the results for HSD may be quite different
from the leading log and leading mass insertion
approximation predictions, making the interpretation of 
BR($\mu\to e\gamma$) in terms of underlying Yukawa couplings
more difficult, but still possible in principle.
We have also shown
that BR($\mu\to e\gamma$) is
independent of $\theta_{13}$, but measurement of this angle
may determine a ratio of sub-dominant Yukawa couplings.

Ultimately one wants
to use the observed neutrino data and constraints (or, if we are lucky,
positive results) from the measurements of the BR($\tau\to\mu\gamma$)
and BR($\mu\to e\gamma$) and other LFV processes to construct
models and discriminate among them. To this end our study will serve
as a framework for such top-down probes, and should help to
establish: (i) if a sequential
dominance mechanism is at work in the neutrino
sector; (ii) if so then what type of dominance (HSD or LSD) is relevant;
(iii) and then (subject to the exceptions discussed in section 3)
the magnitude of the sub-leading neutrino Yukawa couplings
in the flavour basis. 

We emphasise how, within the framework of sequential
dominance, BR($\tau\to\mu\gamma$) and BR($\mu\to e\gamma$) have a
direct interpretation in terms of the underlying Yukawa couplings.
It is worthwhile reiterating, however, that if there are additional sources
of flavour violation, as generically expected in string theory
\cite{graham}, then
these effects will need to be disentangled from the
effects due to the RG running of the see-saw mechanism 
in the CMSSM discused here. Also for simplicity we have neglected 
the effects of phases in a first
global analysis of the CMSSM with a natural neutrino mass hierarchy,
although our results demonstrate that the sign of Yukawa couplings 
can play an important part in determining the mixing angles.

\section*{Acknowledgements}
S.F.K. would like to thank A.Ibarra for useful comments.
T.B. and S.F.K. thank PPARC and CERN for support for this work.

%%%%%%%%%%%%%%%%%%%%%%%%%%%%%%%%%%%%%%%%%%%%%%%%%%%%%%%%%%%%%%%%%%%%%%%
%                                   REFERENCES   
%%%%%%%%%%%%%%%%%%%%%%%%%%%%%%%%%%%%%%%%%%%%%%%%%%%%%%%%%%%%%%%%%%%%%%%%%%%%

\newpage

%\begin{figure}[p]
%\epsfysize=6.5truein
%\epsffile{fig.eps}
%\caption{}
%\label{fig1}
%\end{figure}

\end{document}